\documentclass[aps,twocolumn,amsmath,amssymb,showpacs,floatfix,prl]{revtex4}
\usepackage{graphicx}

\newcommand{\etal}{{\it et al.}}

\begin{document}

\title{The Hall Number in YbRh$_2$Si$_2$}
\author{M. R. Norman}
\affiliation{
Materials Science Division, Argonne National Laboratory, Argonne,
Illinois 60439}
\date{\today}
\begin{abstract}
Recent experimental studies have revealed an abrupt change in the Hall number
at a field tuned quantum critical point in the heavy fermion magnet YbRh$_2$Si$_2$.
We investigate this by calculating the local density band structure for this metal
and the appropriate transport integrals.  We find the Fermi surface
to be multi-sheeted with the two largest sheets having opposite sign contributions
to the Hall number.  Small changes in the f electron occupation are sufficient to reproduce
the observed large change in the Hall number.  This suggests that YbRh$_2$Si$_2$
may be the first example of a quantum critical valence fluctuator.
\end{abstract}
\pacs{75.30.Jb, 71.10.Hf, 72.15.Eb, 71.18.+y}

\maketitle

The nature of the magnetic quantum critical point and the associated non Fermi
liquid behavior in heavy fermion metals is a subject of active debate \cite{GREG,JPCM}.
One of the most interesting suggestions is that at the quantum critical
point, the Fermi surface would transform from a large surface on the paramagnetic
side (with the f electrons participating in the Fermi surface) to a small Fermi surface on
the magnetic side (with the f electrons decoupled from the Fermi surface) \cite{JPCM}.
These authors suggested that the Hall number would be a good test for this scenario.

YbRh$_2$Si$_2$ has emerged as the classic example of a heavy fermion quantum
critical point.  An advantage is that it is stochiometric, and the crystal structure is
relatively simple.  This metal shows pronounced non Fermi liquid behavior in the vicinity of a 
field induced quantum critical point, including a log T divergence of the specific heat coefficient 
which turns into
a power law divergence at sufficiently low temperatures, and a linear T behavior
of the resistivity \cite{YRS1}.  Recently, it has been shown that the Hall number goes through
a significant change at the critical point, suggesting a Fermi surface topology
change as discussed above \cite{HALL}.

The Hall number in the magnetic phase is observed to be around 2 carriers, and in the 
paramagnetic phase around 3 carriers.  This large change (1 carrier per formula
unit) seems to imply that indeed, the f electrons might be decoupling from the Fermi surface.
But there are a number of puzzling issues to address if this is the case.  First, the
ordered magnetic moment is only about 0.002 $\mu_B$ \cite{MOM}, 
which would seem to be inconsistent with localized Yb moments.  Unlike another
small moment case, URu$_2$Si$_2$, there is no evidence for any hidden
order.  In particular, the observed specific heat anomaly at the magnetic transition is
consistent with the small moment.  Moreover, the specific heat coefficient remains large
in the ordered state, which would argue for the existence of heavy fermions in
this phase as well.

In the case of
vanadium doped chromium, there is also a large change in the Hall number at
its quantum critical point \cite{CR02}, which can be simply explained by the
removal of flat parts of the Fermi surface due to magnetic ordering \cite{CR03}.  As
the authors in Ref.~\onlinecite{HALL} remark, the problem with such a scenario
for YbRh$_2$Si$_2$ is the smallness of the ordered moment.  It is difficult to
understand how such a small moment could drive such a large Hall number change.

In fact, very little is known about the ordered phase in YbRh$_2$Si$_2$, except
that it is probably antiferromagnetic in nature.  In particular, the ordering vector is unknown
at this point.  So, as a first step, we will simply consider the
behavior of the Hall number in a paramagnetic calculation.  Surprisingly, we find
that unlike paramagnetic chromium, the Hall number in YbRh$_2$Si$_2$ is a strong
function of the position of the Fermi level.  We then perform a series of calculations
as a function of the position of the f levels.  We find that a small change in the f
level position is sufficient to reproduce the Hall number change.  The topology of
the two Fermi surfaces corresponding to Hall numbers of 2 and 3 are quite
similar, and the f electron occupation in the two cases only differs by 0.03.  This value
is remarkably similar to the 0.03 Rln2 value seen in the entropy associated with
the phase transition on the ordered side \cite{YRS2}.  This implies that the quantum 
critical behavior might
be associated with a small change in valence.  That is, YbRh$_2$Si$_2$ may
be the first example of a quantum critical valence fluctuator.  This scenario
has some similarities to a recent theoretical proposal by P\'{e}pin \cite{PEPIN}.

The calculations in this paper were performed using the local density approximation
within a linear muffin tin orbital scheme \cite{LMTO}.
The exchange-correlation potential used was that of Hedin and Lundqvist \cite{HL}, but
our experience has been that the choice of this potential is not very critical for
f electron systems, as the large Hartree potential of the f electrons dominates the
calculation.  Calculations were performed assuming two choices of muffin tin sphere radii,
and the results were similar.  The results presented in this paper used a sphere radius of
3.886 a.u for Yb, 2.748 a.u. for Rh, and 2.390 a.u for Si (the body centered tetragonal
lattice has an $a$ axis of 4.007$\AA$ and a $c$ axis of 9.858$\AA$ \cite{CRYSTAL}).
These radii were picked based on the various near neighbor separations.
A number of different start configurations were tried, along with various perturbations
during the self-consistent iterations (by shifting the f levels about).  In all cases, the same solution was found.  Results were converged using 120 $k$ points in the irreducible wedge
(1/16th) of the Brillouin zone.  The LMTO code employs the Dirac equation, so spin-orbit
is fully accounted for in the calculation.

After convergence, 1034 $k$ points were generated in the irreducible wedge.  The
eigenvalues were then fit using a Fourier spline series \cite{SPLINE} with 3208 tight
binding functions.  A linear tetrahedron code \cite{LT} was used to
evaluate the density of states and various transport integrals.  The zone was
broken down into 48 $\cdot$ 8$^6$ tetrahedra.  Such a large number of first principles
points and tetrahedra are needed to get a reliable value for the
transport integrals.  This is due to the flat nature of the f electron bands.
For YbRh$_2$Si$_2$, the Hall coefficient was measured in a particular geometry
with the linear response field along the c axis and the current in the plane.
For this geometry, the Hall coefficient in the Boltzmann approximation is given by \cite{ZIMAN}
\begin{equation}
R_H = \sigma_{xyz}/\sigma_{xx}^2
\label{rh}
\end{equation}
where
\begin{equation}
\sigma_{xyz} = \frac{e^3\tau^2}{\hbar\Omega c}\sum_{\vec{k}}
v_x (\vec{v} \times \vec{\nabla}_{k})_z v_y
(-\frac{\partial f}{\partial \epsilon_k})
\label{sigmaxyz}
\end{equation}
\begin{equation}
\sigma_{xx} = \frac{e^2\tau}{\Omega}\sum_{\vec{k}}v_x^2
(-\frac{\partial f}{\partial \epsilon_k})
\label{sigmaxx}
\end{equation}
Here, 1/$\tau$ is the scattering rate,
$\Omega$ the
volume, and $f$ the Fermi distribution function.  The Hall number is the inverse
of $R_H$.  Note that the Boltzmann
approximation should be adequate for our case since
we consider only the T=0 limit.  In particular, the skew scattering correction
to the Hall number vanishes in this limit as demonstrated explicitly
in Ref.~\onlinecite{HALL}.

\begin{figure}
\centerline{
\includegraphics[width=1.7in]{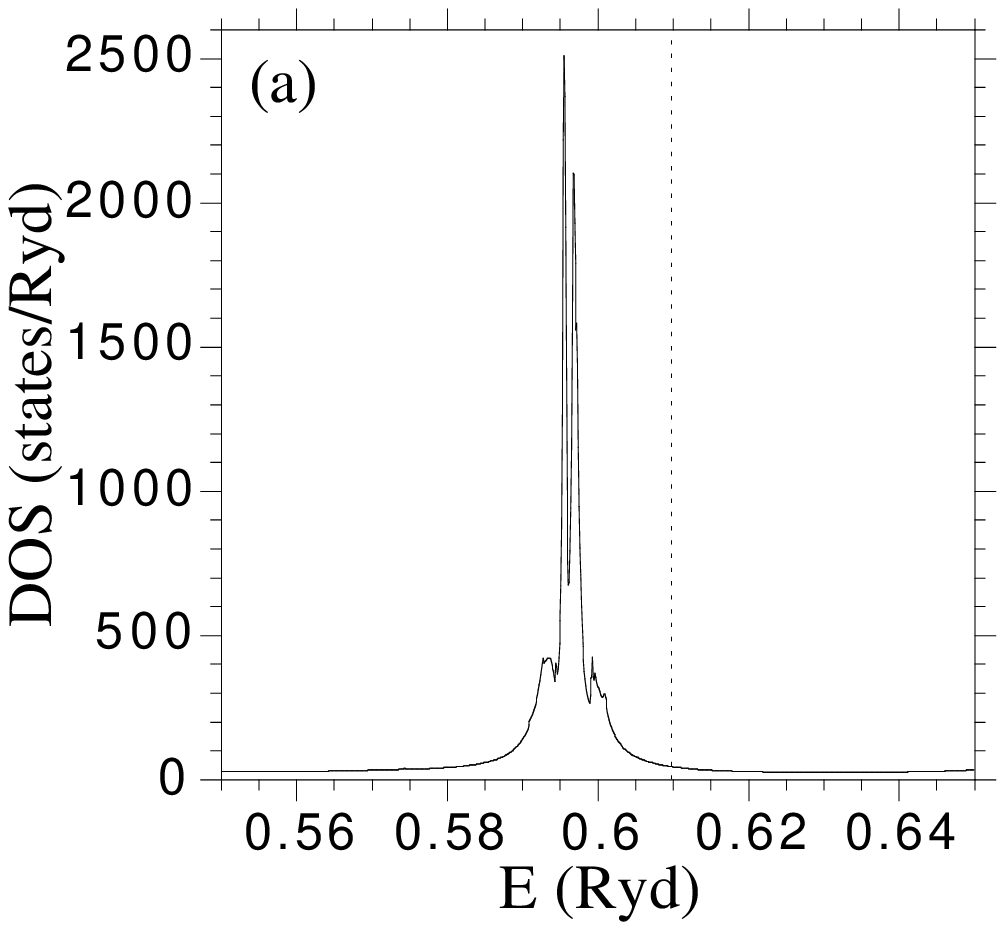}
\includegraphics[width=1.7in]{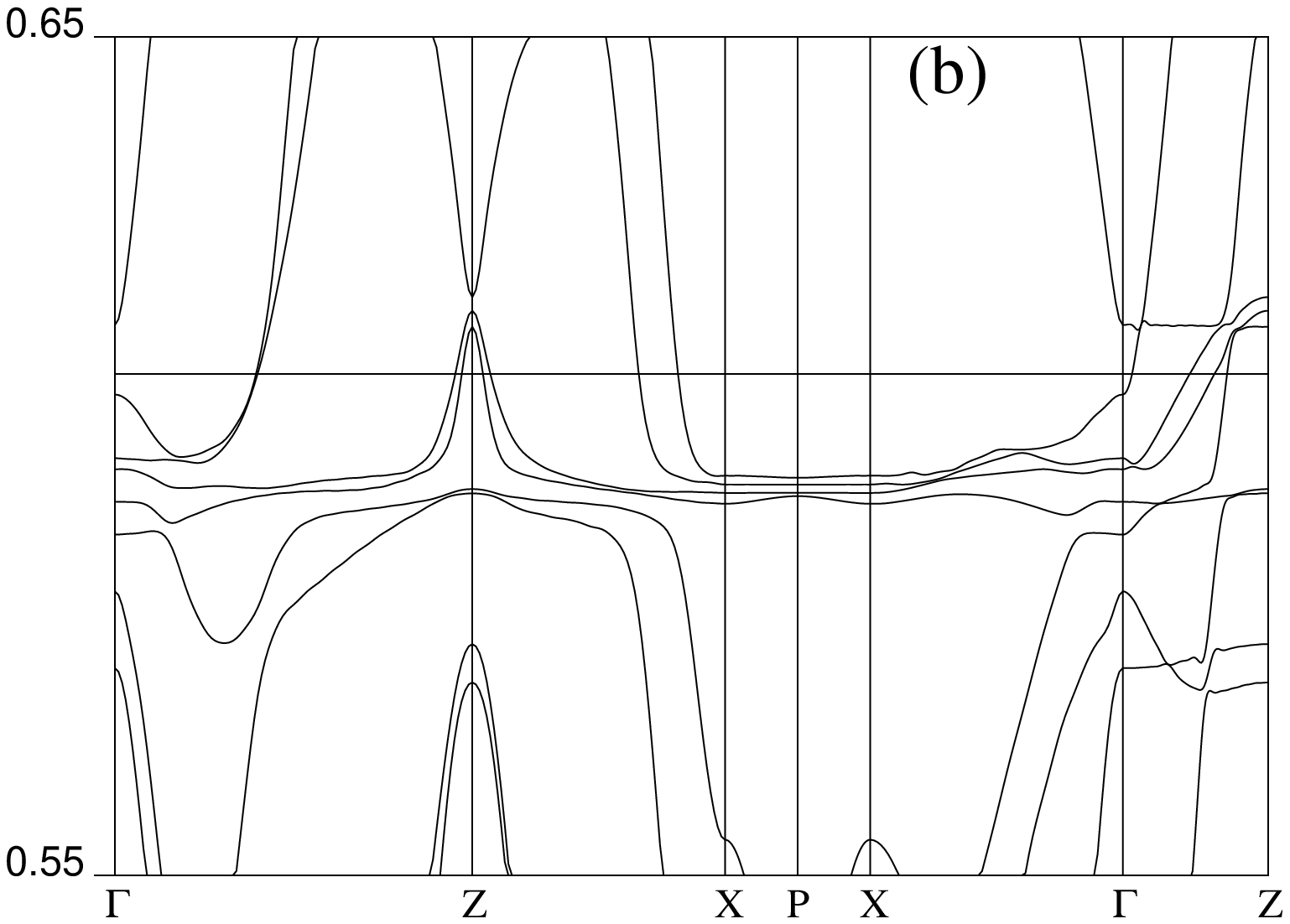}
}
\caption{(a) Density of states and (b) energy bands for YbRh$_2$Si$_2$, with the Fermi
energy denoted by straight lines.  In (b), $Z$ is (002) or (200), $X$ is (110), and $P$ is (111).}
\label{fig1}
\end{figure}

First, we would like to make some general remarks about the electronic structure
of YbRh$_2$Si$_2$.  The spin-orbit splitting is very large (around 1.3 eV), so the
f electron states near the Fermi energy are almost purely of J=7/2 character.
The J=7/2 electron occupation is 7.875.  Some of this charge is due to reanalysis from other sites
(the muffin tin spheres overlap).  To estimate this, we calculate the total J=7/2 occupation
for the lowest 25 bands, and find a value of 8.478.  This implies an f hole value of 0.603,
indicating that
YbRh$_2$Si$_2$ is in the mixed valent regime.  This is a bit of a surprise,
given the heavy fermion nature of this metal.

\begin{figure}
\centerline{
\includegraphics[width=1.7in]{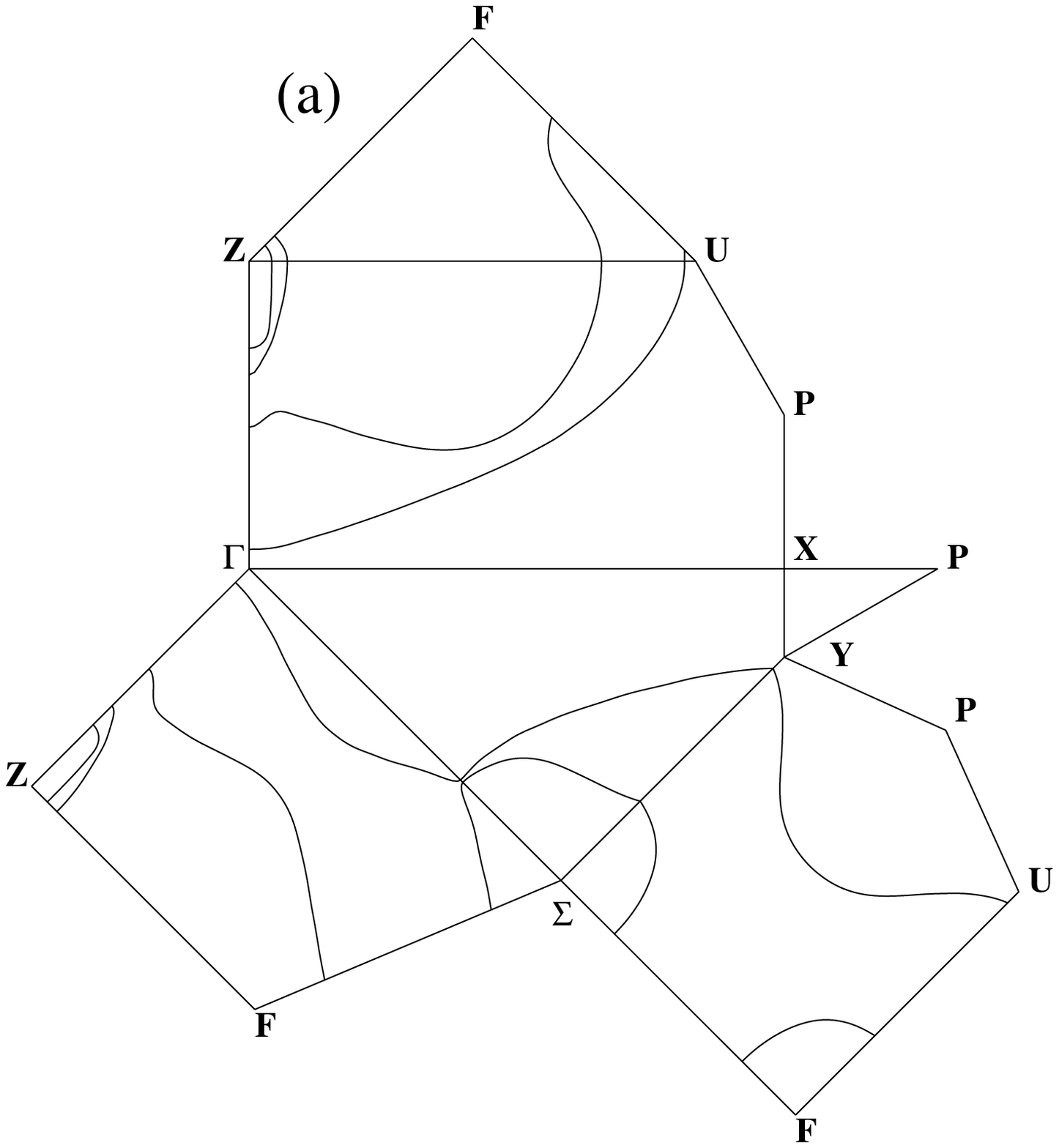}
\includegraphics[width=1.7in]{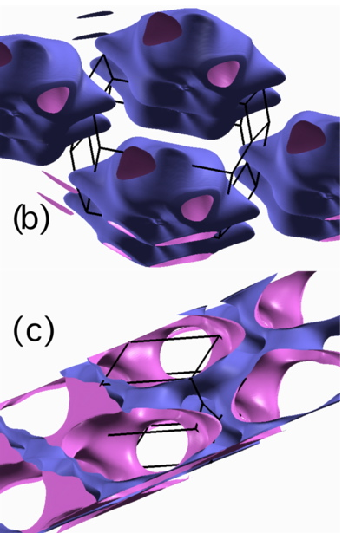}
}
\caption{(a) Fermi surface in symmetry planes of the BCT zone.
$\Sigma$ is ($\tilde{a}$,0,0), $Y$ is ($\tilde{a}$,$\tilde{b}$,0), $F$ is ($\tilde{b}$,0,2), and $U$ 
is ($\tilde{b}$,$\tilde{b}$,2) where $\tilde{a}$=$(c^2+a^2)/c^2$ and
$\tilde{b}$=$(c^2-a^2)/c^2$.  Other zone notation as in Fig.~1.  3D Fermi
surface plot of (b) band 3 and (c) band 4.
The black lines are the BCT zone.}
\label{fig2}
\end{figure}

In Figure 1a, the density of states in the near vicinity of the Fermi energy is shown.  Note
the presence of two very sharp peaks below the Fermi energy.  These
are the local J=7/2 f states which are most flat near the $P$ (111) points of the zone,
as can be seen from Figure 1b.
In the notation of this paper, these states correspond to bands 1-4.  Bands 1 and 2
form small hole ellipsoids around the Z point, as can be seen in Figure 2a.  Band 3 forms
a much larger hole surface centered at the Z point, which resembles a flying saucer
(Figure 2b).  For band 4, the hole surface surrounding the Z point becomes so large,
it forms an interconnected network (Figure 2c).  These two bands have roughly equal
density of states and account for over 97\% of the total density of states.  The J=7/2 average
weight is 62\% for the band 3 Fermi surface, and 50\% for the band 4 Fermi surface
(with an average value of 56\%).
But the total density of states at $E_F$ only corresponds to 8 mJ/mol K$^2$, a rather small
value.  Again, this implies the band result is in the mixed valent regime.

\begin{figure}
\centerline{
\includegraphics[width=3.4in]{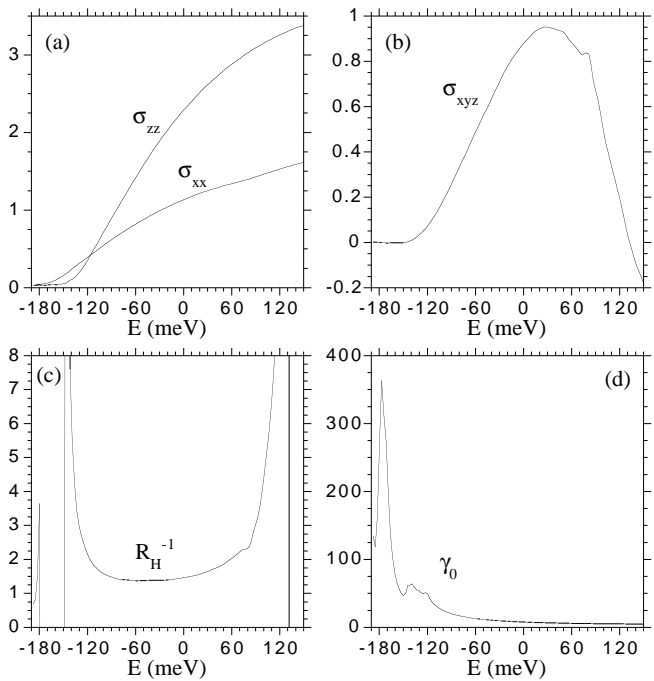}
}
\caption{(a) $\sigma_{ii}$, (b) $\sigma_{xyz}$, (c) $R_H^{-1}$ and (d) density of
states as a function of Fermi level
position relative to $E_F$ (0).}
\label{fig3}
\end{figure}

In Figure 3, we plot $\sigma_{xx}, \sigma_{zz}, \sigma_{xyz},$ and $R_H^{-1}$ as a
function of the Fermi energy.  Unlike the case of paramagnetic chromium, these
quantities are all strong functions of the doping.  In particular, as the Fermi energy approaches
the local f bands, the various $\sigma$ become very small, and the Hall number diverges
rapidly before changing sign.  Interestingly, the Hall number also changes sign for strong
electron doping.  At the Fermi energy, the calculated Hall number is about 1.5, which is
not too far from the zero field value of 2.0 reported in Ref.~\onlinecite{HALL}.  On the other
hand, the $\sigma_{ii}$ integrals at $E_F$ indicate that the residual resistance along the c-axis should
be about half that along the a axis.  Experimentally, this ratio is about 0.9 \cite{YRS2}.
Despite this discrepancy (which could be due to anisotropy of $1/\tau$), this result along
with the experimental resistivity indicates the strong three dimensionality of YbRh$_2$Si$_2$,
which is also consistent with the Fermi surface shown in Figure 2.

Rather than adjusting the Fermi energy (i.e., doping), one can surmise from Figure 3 that
changes of the f level positions will also invoke strong changes in the transport quantities.
From this figure, it is obvious that significant changes can be made by moving the f levels
either towards or further away from the Fermi energy.  We have verified in both cases that
this occurs.  For the purposes of this paper, though, we only show results where the f
levels are moved towards the Fermi energy.  This would seem to be more consistent with
the observed heavy fermion behavior.  Such f level shifts are not unexpected, as they are
often needed to fit deHaas-vanAlphen data on f electron systems, and simply
represent the need to compensate for deficiencies of the local density approximation when
applied to such localized systems.

\begin{figure}
\centerline{
\includegraphics[width=1.7in]{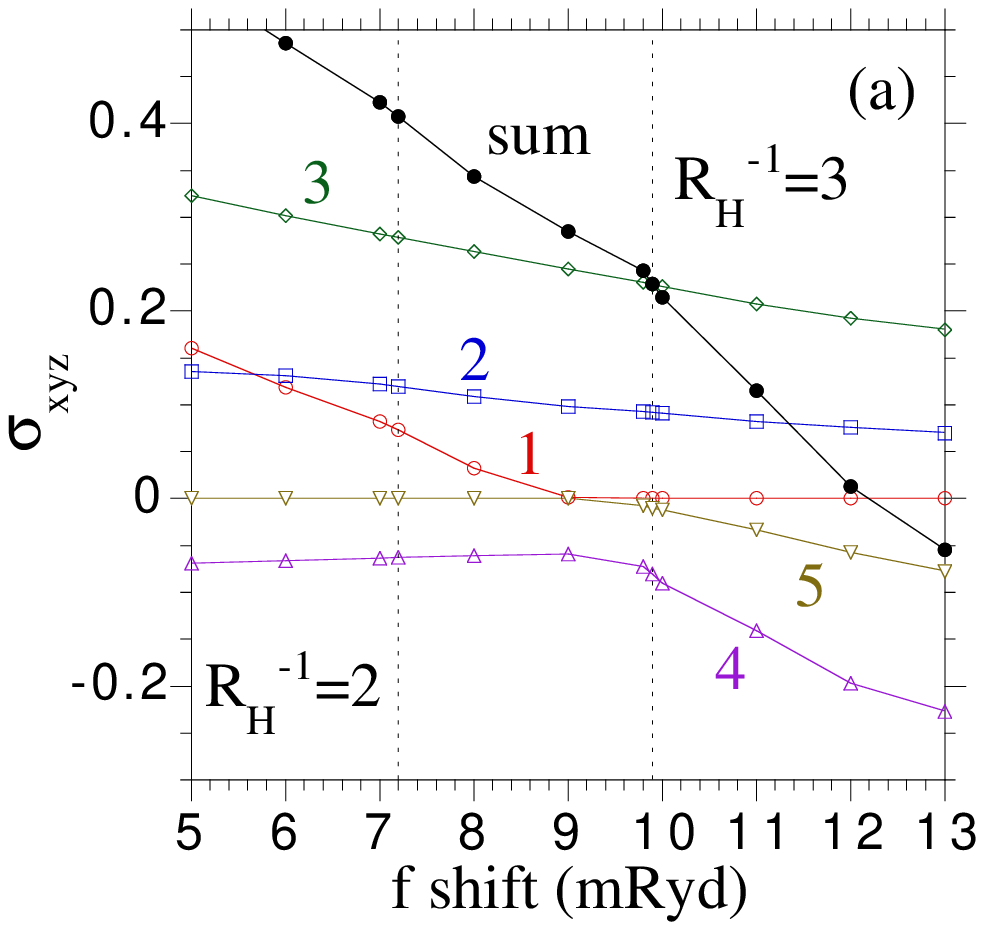}
\includegraphics[width=1.7in]{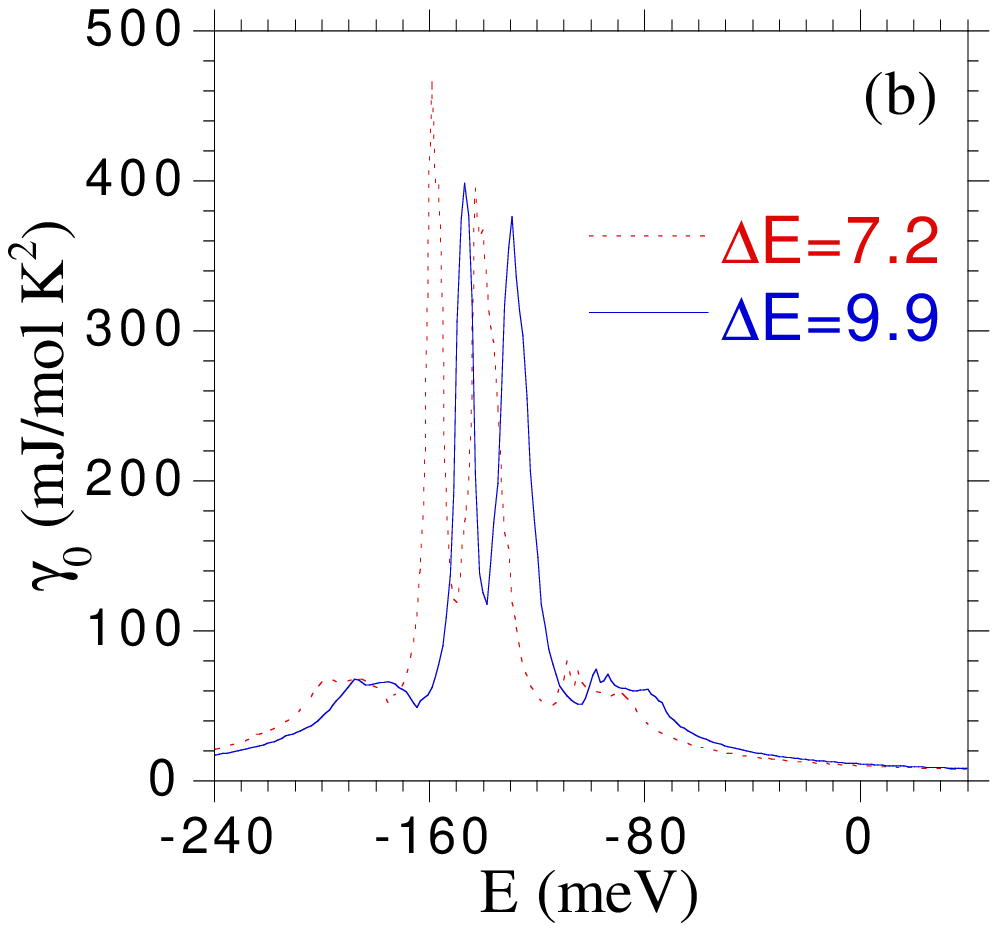}
}
\caption{(a) Band decomposed $\sigma_{xyz}$ and its sum versus f level energy shift.
The dotted lines mark values where the Hall number is either 2 or 3.
(b) Density of states plots for a shift of 7.2 mRyd (Hall number of 2) and 9.9 mRyd
(Hall number of 3).  0 is $E_F$.}
\label{fig4}
\end{figure}

In Figure 4a, we plot the variation of $\sigma_{xyz}$ with respect to the f level shift.
On this plot are dotted lines corresponding to the Hall number values of 2 and 3 found
for the magnetic and paramagnetic phases in Ref.~\onlinecite{HALL}.  As seen here,
bands 1-3 have a positive Hall response as expected (they are hole sheets).  But the
interconnected sheet, band 4, has a negative Hall response (as does band 5, which
appears at higher shifts, and corresponds to a small electron ellipsoid around the 
$\Gamma$ point).  This competition between
a large positive sheet (band 3) and a large negative sheet (band 4) is reminiscent of
chromium \cite{CR03}.

We note that a shift of the f levels by 2.7 mRyd
(37 meV) is sufficient to reproduce the observed Hall number change.  To further
quantify this, we show in Figure 4b the density of states for these two cases.  One
sees that the local f levels relative to $E_F$ shift by only 11 meV.  Given the
large mass renormalization (of order 100 relative to band theory), this would mean
a renormalized energy shift of order  0.1 meV (1 K), a small number indeed.
Moreover, we note from Figures 5a and 5b that the Fermi surfaces for these two cases
are very similar, despite the large change in the Hall number.  As mentioned before,
these two cases differ in f occupation by only 0.03.

\begin{figure}
\centerline{
\includegraphics[width=1.7in]{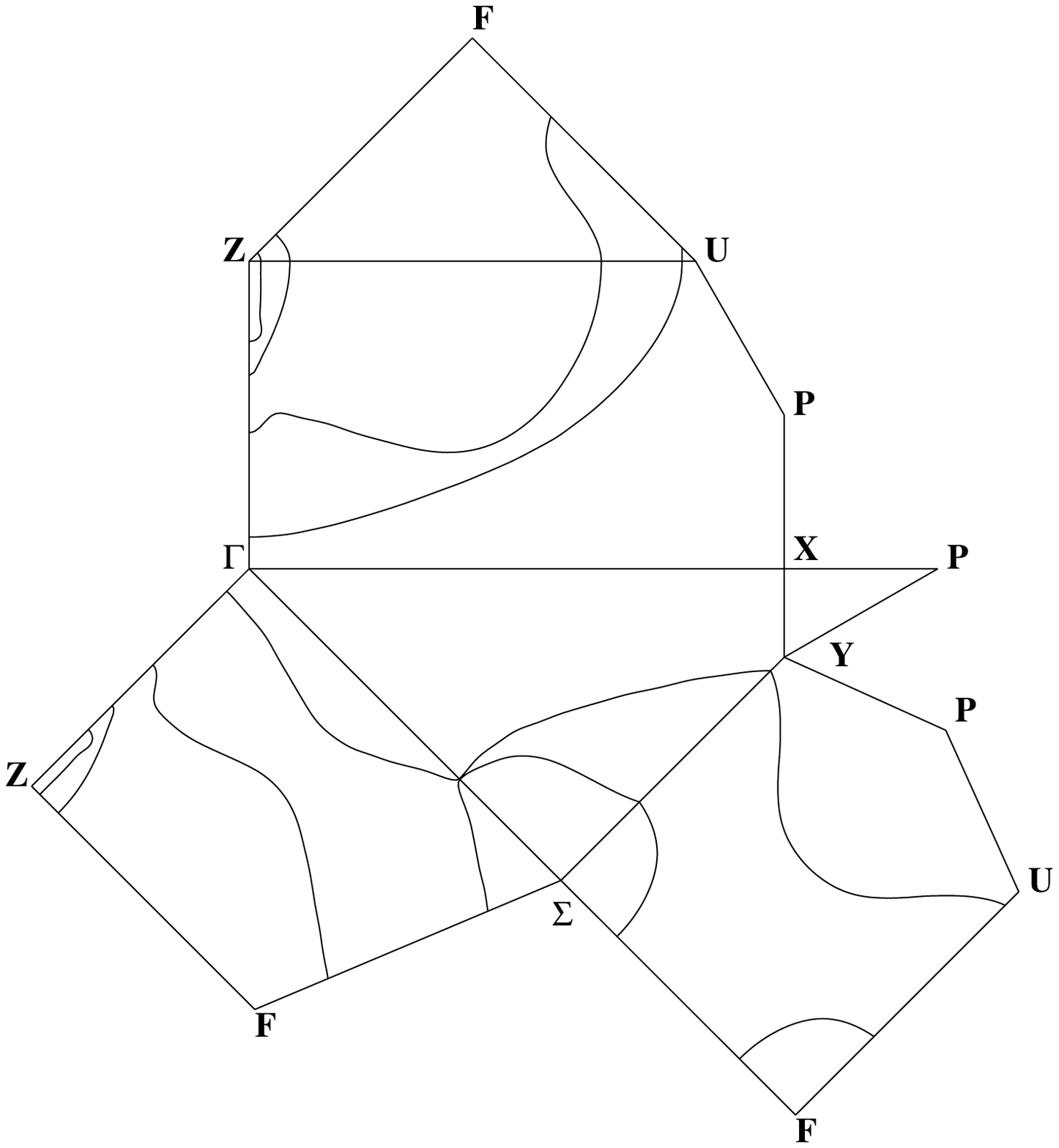}
\includegraphics[width=1.7in]{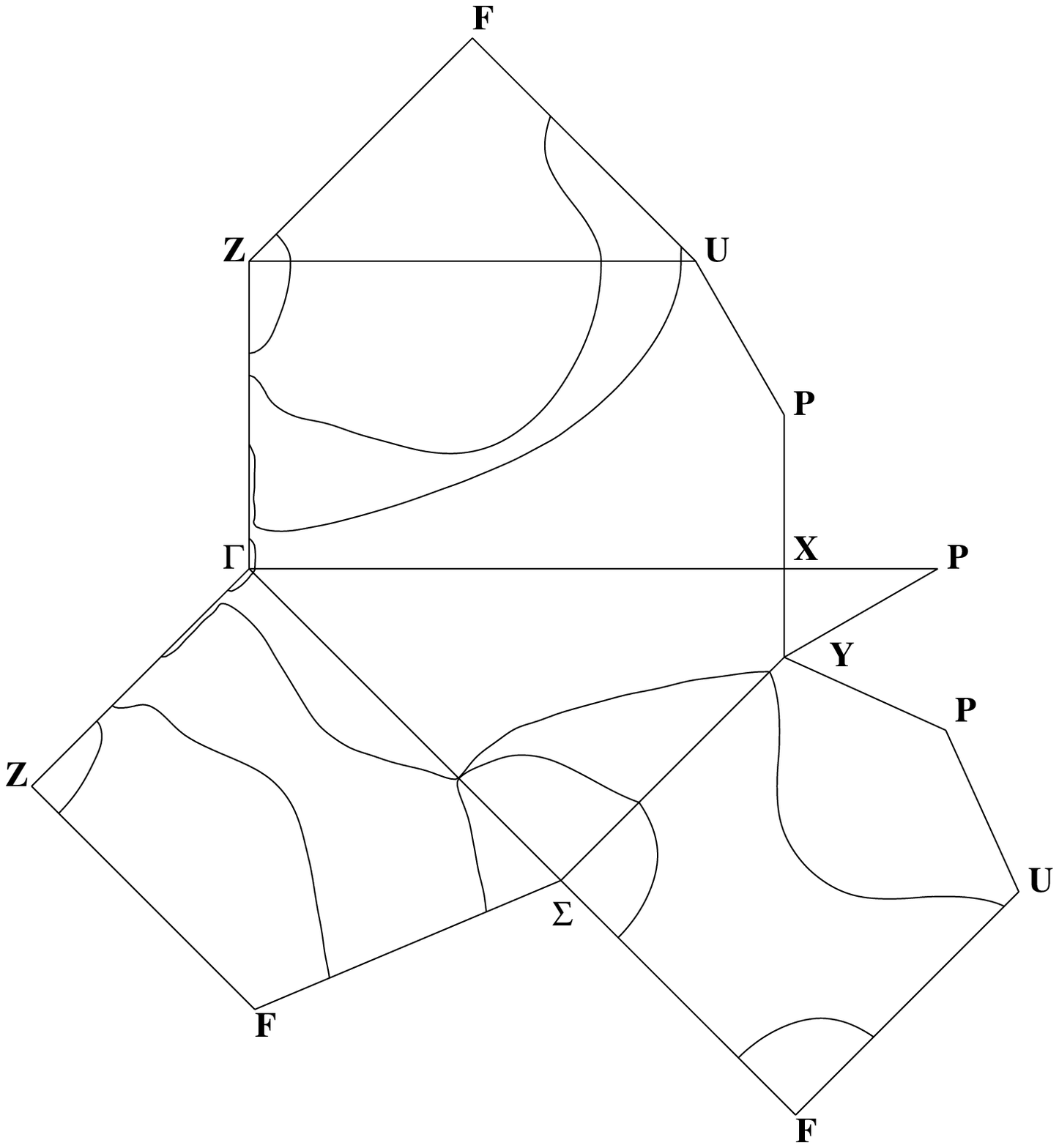}
}
\caption{Fermi surface in symmetry planes of the BCT zone for (a) a Hall number of 2 and
(b) a Hall number of 3.  Zone notation as in Fig.~2.}
\label{fig5}
\end{figure}

In this context, we note that valence changes are common in Yb systems.  The
metal YbInCu$_4$ exhibits a first order valence change at 42 K, with a resulting shift of
the f levels by 50 meV \cite{YIC}.  Perhaps the quantum critical point in YbRh$_2$Si$_2$
is the second order end point for such a transition.  This would seem to be consistent with the
divergence of the Gruneisen ratio seen at the critical point \cite{KUCHLER}.  In this respect,
YbRh$_2$Si$_2$ may be the first example of a quantum critical valence fluctuator.

I would like to thank Catherine P\'{e}pin for discussions, and the hospitality of the KITP where this
work was completed.  The 3D Fermi surface plots were constructed using xcrysden
(http://www.xcrysden.org/).
This work was supported by the U. S. Dept. of Energy, Office of Science,
under Contract No. W-31-109-ENG-38, and in part by the National Science
Foundation under Grant No. PHY99-07949.


\begin{thebibliography}{99}

\bibitem{GREG}
G. R. Stewart, Rev. Mod. Phys. {\bf 73}, 797 (2001).

\bibitem{JPCM}
P. Coleman, C. P\'{e}pin, Q. Si, and R. Ramazashvili, J. Phys. Cond. Matter
{\bf 13}, R723 (2001).

\bibitem{YRS1}
J. Custers \etal, Nature {\bf 424}, 524 (2003).

\bibitem{HALL}
S. Paschen \etal, Nature {\bf 432}, 881 (2004).

\bibitem{MOM}
K. Ishida \etal, Phys. Rev. B {\bf 68}, 184401 (2003).

\bibitem{CR02}
A. Yeh \etal, Nature {\bf 419}, 459 (2002).

\bibitem{CR03}
M. R. Norman, Q. Si, Ya. B. Bazaliy, and R. Ramazashvili, Phys. Rev. Lett. {\bf 90},
116601 (2003).

\bibitem{YRS2}
P. Gegenwart \etal, Acta Phys. Pol. B {\bf 34}, 323 (2003).

\bibitem{PEPIN}
C. P\'{e}pin, Phys. Rev. Lett. {\bf 94}, 066402 (2005).

\bibitem{LMTO}
O. K. Andersen, Phys. Rev. B {\bf 12}, 3060 (1975).

\bibitem{HL}
L. Hedin and B. I. Lundqvist, J. Phys. C {\bf 4}, 2064 (1971).

\bibitem{CRYSTAL}
O. Trovarelli \etal, Phys. Rev. Lett. {\bf 85}, 626 (2000).

\bibitem{SPLINE}
D. D. Koelling and J. H. Wood, J. Comp. Phys. {\bf 67}, 253 (1986).

\bibitem{LT}
G. Lehmann and M. Taut, Phys. Stat. Sol. (b) {\bf 54}, 469 (1972).

\bibitem{ZIMAN}
J. M. Ziman, {\it Electrons and Phonons} (Oxford Univ. Pr., London,
1960), p. 502-503.

\bibitem{YIC}
H. Sato \etal, Phys. Rev. B {\bf 69}, 165101 (2004).

\bibitem{KUCHLER}
R. Kuchler \etal, Phys. Rev. Lett. {\bf 91}, 066405 (2003).

\end{thebibliography}
\end{document}